\documentclass[camera,letterpaper,nomarginnotes,nonarrowgutter]{jpaper}

\usepackage{cite}
\usepackage{amsmath,amssymb,amsfonts}
\usepackage{graphicx}
\usepackage{textcomp}
\usepackage{xcolor}
\usepackage{fancyhdr}
\usepackage[bookmarks=true,breaklinks=true,letterpaper=true,colorlinks,citecolor=blue,linkcolor=blue,urlcolor=blue]{hyperref}
\usepackage{tabularx}
\usepackage{multirow}
\usepackage{listings}
\usepackage{booktabs}
\usepackage{makecell}
\usepackage{enumitem}
\usepackage{balance}
\usepackage{datetime}
\usepackage{datetime2}
\usepackage{setspace}
\usepackage{adjustbox}
\usepackage{siunitx}
\usepackage{glossaries}
\usepackage[linesnumbered]{algorithm2e}
\setkeys{glslink}{hyper=false}

\newcommand{\githubLink}[0]{\url{https://github.com/CMU-SAFARI/Cleaning-up-the-Mess}}

\newcommand{\versionnum}[0]{3.0}

\newcommand{\param}[1]{\textcolor{red}{#1}} % to highlight hardcoded numbers
\newcommand{\xxx}[1]{\param{XXX}} % to highlight hardcoded numbers

\newcommand{\ignore}[1]{}

\newcounter{obs}
\definecolor{nbs}{rgb}{0.88, 0.07, 0.37}
\definecolor{agyc}{rgb}{0.37, 0.88, 0.07}
\definecolor{moegi}{rgb}{0.357, 0.537, 0.188}
\definecolor{burntorange}{rgb}{0.8, 0.33, 0.0}
\definecolor{carmine}{rgb}{0.59, 0.0, 0.09}
\definecolor{ceruleanblue}{rgb}{0.16, 0.32, 0.75}

\newif\ifcamerareadyiterations
\camerareadyiterationsfalse

\newif\ifcameraready
\camerareadytrue

\newif\ifdraft
\draftfalse

\newif\ifblind
\blindtrue

\newif\ifispassdraft
\ispassdraftfalse

\newif\ifshepherd
\shepherdfalse

\newif\ificsdraft
\icsdraftfalse

\newcommand{\omtwo}[1]{#1}

\ificsdraft
\usepackage[colorinlistoftodos,prependcaption,textsize=small]{todonotes}

\paperwidth=\dimexpr \paperwidth + 4cm\relax
\oddsidemargin=\dimexpr\oddsidemargin + 2cm\relax
\evensidemargin=\dimexpr\evensidemargin + 2cm\relax
\marginparwidth=\dimexpr \marginparwidth + 2cm\relax

\newcommand{\nbicscomment}[1]{\todo[size=\scriptsize, linecolor=burntorange, bordercolor=burntorange, backgroundcolor=white!90!burntorange]{\textcolor{burntorange}{#1}}}
\newcommand{\nbics}[1]{\textcolor{burntorange}{#1}}
\else
\newcommand{\nbicscomment}[1]{}
\newcommand{\nbics}[1]{#1}
\fi

\ifcameraready
  \newcommand\nbcr[2]{#2}
  \newcommand{\atbcr}[2]{#2}
  \newcommand{\omcomment}[1]{}
  \newcommand{\nbcrcomment}[1]{}
\else 
    \newcommand\nbcr[2]{\ifnum#1=\value{version}\textcolor{red}{#2}\else{\textcolor{black}{#2}}\fi}
    \newcommand\atbcr[2]{\ifnum#1=\value{version}\textcolor{red}{#2}\else{\textcolor{black}{#2}}\fi}
    \usepackage[colorinlistoftodos,prependcaption,textsize=small]{todonotes}

    \newcommand{\omcomment}[1]{\todo[size=\scriptsize, linecolor=ceruleanblue, bordercolor=ceruleanblue, backgroundcolor=white!90!ceruleanblue]{\textcolor{ceruleanblue}{#1}}}
    \newcommand{\nbcrcomment}[1]{\todo[size=\scriptsize, linecolor=burntorange, bordercolor=burntorange, backgroundcolor=white!90!burntorange]{\textcolor{burntorange}{#1}}}
\fi

\ifshepherd
\usepackage[colorinlistoftodos,prependcaption,textsize=small]{todonotes}

\newcommand{\scomment}[2]{\todo[size=\scriptsize, linecolor=ceruleanblue, bordercolor=ceruleanblue, backgroundcolor=white!90!ceruleanblue]{\textcolor{ceruleanblue}{
\textbf{S#1} #2}}}
\newcommand{\rcomment}[1]{\todo[size=\scriptsize, linecolor=burntorange, bordercolor=burntorange, backgroundcolor=white!90!burntorange]{\textcolor{burntorange}{
\textbf{#1}}}}

\else

\newcommand{\scomment}[2]{}
\newcommand{\rcomment}[1]{}

\fi

\ifdraft
    \usepackage[colorinlistoftodos,prependcaption,textsize=small]{todonotes}
    \newcommand{\nbcomment}[1]{\todo[size=\scriptsize, linecolor=orange, bordercolor=orange, backgroundcolor=white]{\textcolor{nbs}{\textbf{@nb:} #1}}}

    \newcommand{\mma}[1]{\textcolor{orange}{\textbf{[@mma: #1]}}}
    \newcommand\mmatodo[1]{{\textcolor{orange}{#1}}}
    \newcommand{\atbcomment}[1]{\todo[size=\scriptsize, linecolor=orange, bordercolor=orange, backgroundcolor=white]{\textcolor{blue}{\textbf{@atb:} #1}}}

    \newcommand{\omcrcomment}[1]{\todo[size=\scriptsize, linecolor=orange, bordercolor=orange, backgroundcolor=white]{\textcolor{blue}{\textbf{@om:} #1}}}

\else 
    \newcommand{\nbcomment}[1]{}

    \newcommand{\mma}[1]{}
    \newcommand\mmatodo[1]{}
    \newcommand\atbcomment[1]{}
    \newcommand\mmacomment[1]{}

    \newcommand{\omcrcomment}[1]{}
\fi

\ifispassdraft
    \usepackage[colorinlistoftodos,prependcaption,textsize=small]{todonotes}
    \renewcommand{\nbcomment}[1]{\todo[size=\scriptsize, linecolor=orange, bordercolor=orange, backgroundcolor=white]{\textcolor{nbs}{\textbf{@nb:} #1}}}
    
\fi

\definecolor{aliceblue}{rgb}{0.94, 0.97, 1.0}

\newcounter{tkw}
\fancypagestyle{iterationsfirstpage}
{
    \fancyhead{}
    \fancyhead[C]{
    \textcolor{red}{CONFIDENTIAL DRAFT -- DO NOT DISTRIBUTE -- TO APPEAR IN ICS'26} 
    \\
    \emph{\textcolor{blue}{Version \versionnum~---~\today, \ampmtime}}
    }
}

\def\BibTeX{{\rm B\kern-.05em{\sc i\kern-.025em b}\kern-.08em
    T\kern-.1667em\lower.7ex\hbox{E}\kern-.125emX}}
\begin{document}

\newcommand{\affilFirst}{\textsuperscript{$\diamond$}}

\title{{\emph{Extended Abstract:} Re-Evaluating\\the Real-System Modeling Accuracy of Ramulator 2.0}\textsuperscript{$\ast$}}
\author{F. Nisa Bostanc{\i}\affilFirst \quad Haocong Luo\affilFirst \quad Ataberk Olgun\affilFirst \quad Maria Makeenkova \\ Geraldo F. de Oliveira \quad A. Giray Ya{\u{g}}l{\i}k{\c{c}}{\i} \quad Onur Mutlu
\vspace{-3mm}
\\\\
\emph{SAFARI Research Group}
\\
\emph{ETH Z\"urich}}

\maketitle

% \thispagestyle{plain}
% \pagestyle{plain}

%Enables the camera ready header and footer
\ificsdraft
    \thispagestyle{iterationsfirstpage}  
    \pagestyle{plain}
    \pagenumbering{arabic}
\else
    % \renewcommand{\headrulewidth}{0pt}
    % \fancypagestyle{firstpage}{
    %     \fancyhead{} % clear all header and footer fields
    % \renewcommand{\footrulewidth}{0pt}
    % }
    \thispagestyle{plain}
    \pagestyle{plain} 
    \pagenumbering{arabic}
\fi

\renewcommand{\thefootnote}{\fnsymbol{footnote}}
\footnotetext[1]{~This paper is an extended abstract {version} of our full ISPASS~2026 paper~\cite{bostanci2026cleaning}.}
\renewcommand{\thefootnote}{$\diamond$}
\footnotetext{~F. Nisa Bostanc{\i}, Haocong Luo, and Ataberk Olgun are co-primary authors.}
\renewcommand{\thefootnote}{\arabic{footnote}}

\section{{Abstract}}

\begin{figure*}[!t]
    \centering
    \includegraphics[width=1\linewidth]{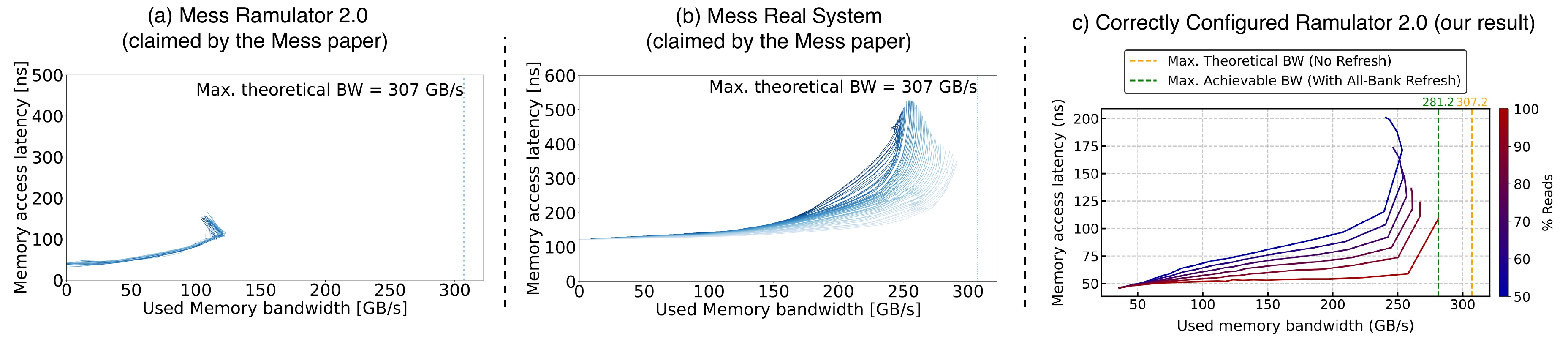}
    \caption{Mess results for Ramulator~2.0 memory system simulation (from
    Fig.~6 in~\cite{mess})~(a), Mess results for an ARM-based real system (from
    Fig.~4 in~\cite{mess})~(b), and our results for Ramulator~2.0 memory system
    simulation after correcting the errors we found in~\cite{mess}~(c).}
    \label{fig:comparison}
\end{figure*}

Cycle-level DRAM simulators {provide accurate and flexible models for DRAM and memory controller operations and enable research on current and future memory systems. Therefore, they are critical} for improving the
performance, efficiency, and robustness of DRAM-based memory systems.  Ramulator~2.0~\cite{ramulator2,ramulator2github} {(successor of Ramulator~\cite{ramulator1,ramulator1github})} is a highly modular and extensible cycle-accurate DRAM simulator that enables rapid exploration of new ideas in DRAM-based memory systems. 

A MICRO 2024 best paper runner-up publication, \emph{A Mess of Memory System
Benchmarking, Simulation and Application Profiling}~\cite{mess}, which we refer
to as ``the Mess paper,'' with all three artifact badges awarded (including
``Reproducible''), proposes a new benchmark to evaluate real and simulated
memory system performance. While doing so, it makes strong negative claims 
about Ramulator~2.0 {and shows unexpected results.} 

{In this talk {and the associated extended abstract}, we demonstrate that these results and claims in the Mess paper} are  incorrect and are due to
configuration and simulator usage errors made in the Mess paper. {We describe four best practices to aid users and developers of simulation tools to avoid such issues in the future.} We emphasize the importance of contacting simulator authors and developers {when unexpected results are observed} (especially {and importantly} before publishing {such results}), 
to ensure these simulators are used with correct configurations and as intended.
Our investigation also aims to stimulate discussion on artifact evaluation practices and on mechanisms for correcting results and artifacts after publication.
To aid future works and reproduction of all our results, we open source all our code and scripts at \githubLink{}.
We
refer the reader to our full ISPASS~2026 paper~\cite{bostanci2026cleaning} {and its artifact~\cite{cutmess.github}} for
the complete analysis, detailed methodology, and extended results.

\section{{Summary of Our Analysis of the Evaluation of Ramulator~2.0 in~\cite{mess}}}
\label{sec:errors}

\subsection{Mess Paper's Ramulator~2.0 Results}

Fig.~\ref{fig:comparison}-(a) and (b) (copied from Fig.~6 and Fig.~4
in~\cite{mess}) show average memory latency as a function of throughput (used
memory bandwidth) {presented in~\cite{mess}} for the Mess benchmark using a Ramulator~2.0 simulation and a
real system, respectively.\footnote{The Mess benchmark runs two concurrent
workloads with \emph{Stream} and \emph{Pointer-Chase} memory access patterns, respectively. The interval between subsequent memory requests in Stream is
controlled using NOPs to vary memory intensity. Each
bandwidth--latency curve depicts system performance results in increasing memory intensity order, 
for a read/write mixture in the Stream workload ranging from 50\% to 100\% reads.} From
these two figures, the Mess paper concludes that Ramulator~2.0 ``poorly resembles
the actual system performance'' {(Section I of~\cite{mess})} and writes ``The simulated memory latency is
unrealistically low and the maximum simulated memory bandwidth is only 126 GB/s
which is less than a half of the 292 GB/s measured in the actual system''
{(Section IV of~\cite{mess})}. The paper attributes this discrepancy between \omtwo{Mess} benchmark results of Ramulator~2.0
simulations and the real system to Ramulator~2.0-induced simulation
errors. 

\subsection{Our Analysis} 
We carefully investigate the Mess paper's open source
artifacts~\cite{messgit,messzenodo} and have communicated with its authors to identify
the root causes of these errors. 
{We identify various issues in three categories.}

\noindent\textbf{Reproducibility Issues.}
{We note that it {was} 
{\textit{not} possible}
to reproduce {\emph{any}} Ramulator~2.0 results
from the {Mess} paper artifacts available at the time of {the} publication and artifact
evaluation {of the {Mess} paper}~\cite{messgit,messzenodo}. Thus, we contacted the 
authors {of~\cite{mess} on 16 October 2024} to obtain
1)~the Ramulator~2.0 source code and 2)~the {Mess} benchmark memory traces used in
Ramulator~2.0 simulations reported in~\cite{mess}. The missing sources
{were} {eventually} uploaded to {the} {Mess} GitHub
repository~\cite{messgit} {on 18 July 2025},
\emph{10 months after} the {Mess}
paper {was} awarded {(on 8 September 2024)} all three artifact evaluation badges.}

\noindent\textbf{Fundamental Configuration Errors.} 
{We identify two fundamental configuration errors in the Mess paper~\cite{mess}'s evaluation of
Ramulator~2.0, which result in the low memory bandwidth and low memory latency that the 
paper attributes to Ramulator~2.0's simulation inaccuracy.}

First, the Mess paper's
scripts instantiate a single-channel DDR5 Ramulator~2.0 memory system and
extrapolate single-channel bandwidth to 16-channel bandwidth by multiplying the
observed bandwidth by 8 instead of 16. This is because the Mess paper
incorrectly assumes that a DDR5 channel in Ramulator~2.0 is 64 bits wide,
consisting of two independent 32-bit channels. However, a DDR5 channel is 32
bits wide in Ramulator~2.0,\footnote{The \texttt{m\_channel\_width} parameter in
\texttt{DDR5.cpp} in the Ramulator~2.0 repository~\cite{ramulator2github}.}
which is consistent with JEDEC DDR5 standards~\cite{jedecddr5c,jedecddr5udimm,jedecddr5rdimm}.
{As a result,} the Mess paper {compares an (extrapolated)} 8-channel {memory system in Ramulator~2.0 to}
{a} 16-channel {real memory system}, {and thus, observes a lower memory bandwidth.}

Second, the Mess
paper's scripts {run} Ramulator~2.0 with unrealistically
low latency configurations: the cache latency is set to \emph{0} CPU clock
cycles and each CPU core has 1024 miss-status holding registers. Such a
configuration has significantly smaller core-to-DRAM round-trip latency and
significantly lower queuing delay than a real CPU {leading to} {a lower overall memory latency. This unrealistically low configuration is in part responsible for the low latency observed in~\cite{mess}.}

\sloppy
\noindent\textbf{Scholarship Issues.} 
{We identify that the Mess paper does not disclose 1)~the unrealistically low latency configurations ({which the Mess \nbics{artifacts} hardcoded} via changing the source code of Ramulator~2.0) and 2)~differences between the methodologies of evaluating real and simulated memory systems ({which we} identified via email exchanges with the Mess paper's authors).}

\section{Our {Major} Results}
\label{sec:results}

\subsection{Evaluation Methodology} 
To show that Ramulator~2.0, when configured
properly and used correctly, yields realistic bandwidth--latency curves {that closely resemble a real system}, we
introduce a new Ramulator~2.0 frontend, the \emph{Mess Request Generator}, which
emulates the Mess benchmark memory access patterns by generating a configurable
mixture of random (pointer-chasing) and stream read/write requests with a
configurable NOP frequency to modulate system load. We evaluate a 16-channel
(32-bit each) DDR5-4800AN system, {sweeping read/write requests in the Stream workload from 50\% to 100\% reads} and sweeping NOP
frequency from 1 to 10000.

\subsection{Ramulator~2.0 Key Results} 
Fig.~\ref{fig:comparison}-(c) shows the bandwidth--latency curves generated using 
Ramulator~2.0 with the Mess Request Generator frontend. The maximum theoretical 
bandwidth for 16 channels of DDR5-4800 (indicated by the orange line in the figure)
is 307.2 GB/s ($16\times{}$19.2 GB/s). 
{Accounting for
all-bank periodic refresh operations that delay data
transfers,} the \emph{maximum achievable bandwidth} (indicated by the green 
line in the figure) is calculated as 281.2 GB/s {(following the equation in \S4 
in~\cite{bostanci2026cleaning}).}

We make three key observations.
First, the maximum achieved bandwidth (281.1 GB/s for {a read ratio of 1, i.e.,}
100\% reads, and NOP {frequency of 1, i.e., no NOP cycles}) almost exactly matches
the maximum achievable bandwidth (281.2 GB/s). 
Second, as the fraction of write memory requests increases, the \nbcr{5}{bandwidth--latency} 
curves are curved more aggressively
towards the top left corner of the figure because a larger fraction of
write memory requests induces a higher latency and bandwidth penalty
due to more frequently incurred DRAM bus turnaround 
latency~\cite{lee2010dram}.\footnote{{This phenomenon is explained in more detail in~\cite{bostanci2026cleaning}.}} 
Third, the latency of random read requests 
increases with used memory bandwidth, because interference from Stream requests
in the memory controller queues increases the queuing delay for random reads. We
conclude that Ramulator~2.0, when configured and used correctly, yields
reasonable and realistic bandwidth--latency curves that resemble those obtained from real
systems by~\cite{mess} {(see Section 3, Fig. 3 in~\cite{mess})}. The results {we present in this section}
{thus} show that the Ramulator~2.0 results presented in~\cite{mess} are {incorrect} and are \emph{not}
representative of Ramulator~2.0's modeling accuracy {and capability}.

\section{{Best Practices}}
To help users and developers avoid such errors {we identified in \nbics{the Mess paper}~\cite{mess}}, we recommend four best practices.

{First, we recommend Ramulator~2.0 users familiarize themselves 
with JEDEC standard documents (e.g., the JEDEC DDR5 {DRAM} standard~\cite{jedecddr5c}) and the simulator documentation. Colloquial technical terms
may differ from the precise definitions in official standards. Thus, Ramulator~2.0 implements the DRAM standards based on the JEDEC standard documents. Therefore, the users {of Ramulator~2.0} should be aware of the differences between colloquial technical terms and standard definitions, and configure simulators based on the simulator implementation.}

{Second, users should} understand Ramulator~2.0's memory organization and configurations, which are
documented in the code and the repository. {The first Ramulator~2.0 configuration error in the Mess paper~\cite{mess} could have been avoided with the first two best practices.}

{Third, users should} understand the simulation
methodology \nbics{and the semantics of reported simulator statistics} sufficiently to use the correct statistics.
{The third best practice could have {prevented}
{simulator usage errors in~\cite{mess} (explained in detail in~\cite{bostanci2026cleaning})}.}

{Fourth, users should} communicate with the simulator developers or user community when unexpected results are observed {to ensure correct simulator usage}. {Questions about a simulator and open source software should be handled in the source repository (i.e., the GitHub repository~\cite{ramulator2github} for Ramulator~2.0) via discussion with developers and the user community, in true open source spirit~\cite{lakhani2003open}.} {The first three best practices each prevent a specific class of error. The fourth best practice, however, could \nbics{by itself} have prevented all configuration and simulator usage errors in~\cite{mess}.}

\section{Conclusion}
We demonstrated that the claims made by the Mess paper~\cite{mess} about
Ramulator~2.0~\cite{ramulator2} 
are incorrect and are due
to simulator configuration errors made in the Mess paper.
We strongly encourage the computer
architecture community to consider our corrections to prevent the propagation of
inaccurate results and to maintain the reliability of the scientific record.
More broadly, simulators should be configured, understood, and used carefully
and rigorously before making claims about them being wrong, and the authors of
freely available simulators should be properly contacted before publishing
strong claims about such simulators. 
{This is especially important for freely available open source research software.} 
To
facilitate transparency and reproducibility, we open source all our code and
results at \githubLink{}. The full version of this work, including the complete
methodology and extended results, appears in our ISPASS~2026
paper~\cite{bostanci2026cleaning} {and also its extended version on arXiv~\cite{cutmessarxiv}}. {We also conduct scientific tutorials on Ramulator in major conferences, including ASPLOS 2026~\cite{rndbasplos26,rndbasplos26live}, ISCA 2026~\cite{rndbisca26}, and ICS 2026~\cite{rndbics26}, whose materials are freely available online to the entire world.}

\section*{Acknowledgments}
% \begin{acks}
{The full version of this work, including the complete
methodology and extended results, appears in our ISPASS~2026
paper~\cite{bostanci2026cleaning} and also its extended version on arXiv~\cite{cutmessarxiv}}.
{We thank the anonymous reviewers of {ISPASS 2026} for feedback. 
{The anonymous reviewers heavily scrutinized this paper and provided
numerous suggestions, {in some cases} sentence by sentence, to ensure that each statement 
in this paper is correct and firmly backed up by data. If the strict scrutiny applied to this paper 
had been consistent across the scientific review process, 
we believe the Mess paper would \emph{not} have been published in its current state.}
We believe our results and analyses in this paper open up broader {questions regarding} the review and artifact {evaluation} processes.
{In particular, it is critical for tool papers that claim incorrect results in freely available open-source software to be held to the standard of being asked to have checked with the authors and maintainers of the software first, to establish the accuracy of their results.}

We thank practitioners from 
{many major companies in industry and researchers {at} many institutions in academia}
who {have been using} Ramulator and {providing} valuable feedback 
since the release of the first {version in 2015, as well as the second version in 2023}. Their experiences and comments helped {us} validate the {practicality, usability, and accuracy}
of our tools.
{Some of these folks have also read the Mess paper and expressed their surprise that it was selected as a "best paper runner-up" and even accepted, even though its authors had no communication with the developers and maintainers of Ramulator~2.0.}
{We thank the} SAFARI Research Group members for
{constructive} feedback and the stimulating {\& open} intellectual, {scientific} {and scholarly} {environment {they provide}.}
We acknowledge the generous gift funding provided by our industrial partners
({especially} Google, Huawei, Intel, Microsoft), which has been instrumental in
enabling the {extensive} research we have been conducting on memory systems \nbcr{5}{for more than two decades}~\cite{kim2014flipping,mutlu2025modern,mutlu2013memory,mutlu2025memory,mutlu2019rowhammer,mutlu2023fundamentally,mutlu2019processing,cai2017flashtbd,mutlu2014research,singh2021fpga,mutlu2017rowhammer,mutlu2020intelligentdate,oliveira2022accelerating,mutlu2024memory,mutlu2023experimentalretrospective,mutlu2023raidrretrospective,mutlu2023rowhammerretrospective, mutlu2023tesseractretrospective, mutlu2023selfretrospective,luo2026ramulator2.1,yuksel2026memory,olgun2026drambender,oliveira2021damov,mutlu2015main,gomez2021benchmarking,kakolyris2026columnkeeper,ghose2019processing}. This work was in part
supported by {a} Google Security and Privacy Research Award and the Microsoft
Swiss Joint Research Center.}
% \end{acks}

\balance
\bibliographystyle{unsrt}

\bibliography{refs}

@ARTICLE{ramulator1,
  title = {{Ramulator: A Fast and Extensible DRAM Simulator}},
  author = {Kim, Yoongu and Yang, Weikun and Mutlu, Onur},
  journal = {{IEEE CAL}},
  year = {2015},
}

@ARTICLE{ramulator2,
  author={Luo, Haocong and Tugrul, Yahya Can and Bostanci, F. Nisa and Olgun, Ataberk and Yaglikci, A. Giray and Mutlu, Onur},
  journal={IEEE CAL}, 
  title={{Ramulator 2.0: A Modern, Modular, and Extensible DRAM Simulator}}, 
  year={2024}}

@INPROCEEDINGS{mess,
  author={Esmaili-Dokht, Pouya and Sgherzi, Francesco and Girelli, Valéria Soldera and Boixaderas, Isaac and Carmin, Mariana and Monemi, Alireza and Armejach, Adrià and Mercadal, Estanislao and Llort, Germán and Radojković, Petar and Moreto, Miquel and Giménez, Judit and Martorell, Xavier and Ayguadé, Eduard and Labarta, Jesus and Confalonieri, Emanuele and Dubey, Rishabh and Adlard, Jason},
  booktitle={MICRO}, 
  title={{A Mess of Memory System Benchmarking, Simulation and Application Profiling}}, 
  year={2024}}

@misc{messgit,
  title={{Mess benchmark}},
  author={{Memory systems for HPC and AI @BSC}},
  year={2024},
  howpublished={\url{https://github.com/bsc-mem/Mess-benchmark}}
}

@misc{messzenodo,
  title={{A Mess of Memory System Benchmarking, Simulation and Application Profiling}},
  author={{Esmaili-Dokht, Pouya}},
  year={2024},
  howpublished={\url{https://zenodo.org/records/13748674}}
}

@article{oliveira2021damov,
  title={{DAMOV: A New Methodology and Benchmark Suite for Evaluating Data Movement Bottlenecks}},
  author={Oliveira, Geraldo F and G{\'o}mez-Luna, Juan and Orosa, Lois and Ghose, Saugata and Vijaykumar, Nandita and Fernandez, Ivan and Sadrosadati, Mohammad and Mutlu, Onur},
  journal={IEEE Access},
  year={2021},
}

@manual{jedecddr5c,
  title="{JESD79-5C: DDR5 SDRAM Standard}",
  author="{{JEDEC}}",
  year={2024}
}

@manual{jedecddr5udimm,
  title={{JESD308B: DDR5 Unbuffered Dual Inline Memory Module (UDIMM) Common Standard}},
  author={{JEDEC}},
  year={2025}
}

@manual{jedecddr5rdimm,
  title={{JESD305A: DDR5 Registered Dual Inline Memory Module (RDIMM) Common Standard}},
  author={{JEDEC}},
  year={2025}
}

@misc{ramulator1github,
	title        = {{Ramulator}},
	author       = {SAFARI Research Group},
	howpublished = {\url{https://github.com/CMU-SAFARI/ramulator}},
    year         = {2015}
}

@misc{ramulator2github,
	title        = {{Ramulator 2.0}},
	author       = {SAFARI Research Group},
	howpublished = {\url{https://github.com/CMU-SAFARI/ramulator2}},
    year         = {2023}
}

@misc{lee2010dram,
  title={{DRAM-Aware Last-Level Cache Writeback: Reducing Write-Caused Interference in Memory Systems}},
  author={Lee, Chang Joo and Narasiman, Veynu and Ebrahimi, Eiman and Mutlu, Onur and Patt, Yale N},
  year={2010},
  howpublished={Technical Report HPS-2010-002}
}

@article{lakhani2003open,
  title={{How Open Source Software Works: ``Free'' User-to-user Assistance}},
  author={Lakhani, Karim R and Von Hippel, Eric},
  journal={Research Policy},
  year={2003}
}

@misc{cutmess.github,
	title        = {{Cleaning up the Mess Source Code}},
	author       = {SAFARI Research Group},
	howpublished = {\url{https://github.com/CMU-SAFARI/Cleaning-up-the-Mess}},
    year         = {2026}
}

@INPROCEEDINGS{bostanci2026cleaning,
  author={Bostanci, F. Nisa and Luo, Haocong and Olgun, Ataberk and Makeenkova, Maria and Oliveira, Geraldo F. and Yaglikci, A. Giray and Mutlu, Onur},
  booktitle={ISPASS},
  title={{Cleaning up the Mess: Re-Evaluating the Real-System Modeling Accuracy of Ramulator 2.0}},
  year={2026}}

@misc{mutlu2025modern,
	title        = {{A Modern Primer on Processing in Memory}},
	author       = {Mutlu, Onur and Ghose, Saugata and G{\'o}mez-Luna, Juan and Ausavarungnirun, Rachata},
	year         = 2025,
	howpublished = {arXiv:2012.03112 [cs.AR]}
}

@inproceedings{mutlu2013memory,
	title        = {{Memory Scaling: A Systems Architecture Perspective}},
	author       = {Mutlu, Onur},
	year         = 2013,
	booktitle    = {IMW}
}

@inproceedings{mutlu2025memory,
  title={{Memory-Centric Computing: Solving Computing’s Memory Problem}},
  author={Mutlu, Onur and Olgun, Ataberk and Yuksel, Ismail Emir},
  booktitle={IMW},
  year={2025}
}

@article{mutlu2019rowhammer,
	title        = {{RowHammer: A Retrospective}},
	author       = {Mutlu, Onur and Kim, Jeremie S},
	year         = 2019,
	journal      = {TCAD},
	booktitle    = {COSADE}
}

@inproceedings{mutlu2023fundamentally,
	title        = {{Fundamentally Understanding and Solving RowHammer}},
	author       = {Mutlu, Onur and Olgun, Ataberk and Yaglikci, A. Giray},
	year         = 2023,
	booktitle    = {ASP-DAC}
}

@inproceedings{mutlu2019processing,
	title        = {{Processing Data Where It Makes Sense: Enabling In-Memory Computation}},
	author       = {Mutlu, Onur and Ghose, Saugata and G{\'{o}}mez-Luna, Juan and Ausavarungnirun, Rachata},
	year         = 2019,
	booktitle    = {MICPRO}
}

@inproceedings{mutlu2024memory,
  title={{Memory-Centric Computing: Recent Advances in Processing-in-DRAM}},
  author={Mutlu, Onur and Olgun, Ataberk and Oliveira, Geraldo F and Yuksel, Ismail E},
  booktitle={IEDM},
  year={2024}
}

@article{oliveira2022accelerating,
	title        = {{Accelerating Neural Network Inference with Processing-in-DRAM: From the Edge to the Cloud}},
	author       = {Oliveira, Geraldo F and G{\'o}mez-Luna, Juan and Ghose, Saugata and Boroumand, Amirali and Mutlu, Onur},
	year         = 2022,
	journal      = {IEEE Micro}
}

@inproceedings{mutlu2020intelligentdate,
	title        = {{Intelligent Architectures for Intelligent Computing Systems}},
	author       = {Mutlu, Onur},
	year         = 2021,
	booktitle    = {DATE}
}

@inproceedings{mutlu2017rowhammer,
	title        = {{The RowHammer Problem and Other Issues We May Face as Memory Becomes Denser}},
	author       = {Mutlu, Onur},
	year         = 2017,
	booktitle    = {DATE}
}

@article{singh2021fpga,
	title        = {{FPGA-Based Near-Memory Acceleration of Modern Data-Intensive Applications}},
	author       = {Singh, Gagandeep and Alser, Mohammed and Cali, Damla Senol and Diamantopoulos, Dionysios and G{\'o}mez-Luna, Juan and Corporaal, Henk and Mutlu, Onur},
	year         = 2021,
	journal      = {IEEE Micro}
}

@article{mutlu2014research,
	title        = {{Research Problems and Opportunities in Memory Systems}},
	author       = {Mutlu, Onur and Subramanian, Lavanya},
	year         = 2014,
	journal      = {SUPERFRI}
}

@article{cai2017flashtbd,
	title        = {{Error Characterization, Mitigation, and Recovery in Flash Memory Based Solid-State Drives}},
	author       = {Y. Cai and S. Ghose and E. F. Haratsch and Y. Luo and O. Mutlu},
	year         = 2017,
	journal      = {Proc. IEEE}
}

@misc{cutmessarxiv,
      title={{Cleaning up the Mess: Re-Evaluating the Real-System Modeling Accuracy of Ramulator 2.0}}, 
      author={F. Nisa Bostanci and Haocong Luo and Ataberk Olgun and Maria Makeenkova and Geraldo F. Oliveira and A. Giray Yaglikci and Onur Mutlu},
      year={2026},
      eprint={2510.15744},
      archivePrefix={arXiv},
      primaryClass={cs.AR},
      url={https://arxiv.org/abs/2510.15744},
      howpublished={arXiv:2510.15744 [cs.AR]}
}

@misc{rndbasplos26,
	title        = {{The First Tutorial on Ramulator and DRAM Bender: Cutting-Edge Infrastructures for Real and Future Memory System Evaluation at ASPLOS'26}},
	author       = {F. Nisa Bostanci and Ataberk Olgun and Ismail Emir Yuksel and Haocong Luo and Onur Mutlu},
	howpublished = {\url{https://events.safari.ethz.ch/asplos26-ramulator-drambender/}},
    year         = {2026}
}

@misc{rndbisca26,
	title        = {{The Second Tutorial on Ramulator and DRAM Bender: Cutting-Edge Infrastructures for Real and Future Memory System Evaluation at ISCA'26}},
	author       = {F. Nisa Bostanci and Ataberk Olgun and Ismail Emir Yuksel and Haocong Luo and Onur Mutlu},
	howpublished = {\url{https://events.safari.ethz.ch/isca26-ramulator-drambender/}},
    year         = {2026}
}

@misc{rndbics26,
	title        = {{The Third Tutorial on Ramulator and DRAM Bender: Cutting-Edge Infrastructures for Real and Future Memory System Evaluation at ICS'26}},
	author       = {F. Nisa Bostanci and Ataberk Olgun and Ismail Emir Yuksel and Haocong Luo and Onur Mutlu},
	howpublished = {\url{https://events.safari.ethz.ch/ics26-ramulator-drambender/}},
    year         = {2026}
}

@misc{rndbasplos26live,
	title        = {{The First Tutorial on Ramulator and DRAM Bender @ ASPLOS'26: Livestream}},
	author       = {F. Nisa Bostanci and Ataberk Olgun and Ismail Emir Yuksel and Haocong Luo and Onur Mutlu},
	howpublished = {\url{https://www.youtube.com/live/6uAfgKs1JKI}},
    year         = {2026}
}

@misc{mutlu2023tesseractretrospective,
  title={{Retrospective: A Scalable Processing-in-Memory Accelerator for Parallel Graph Processing}},
  author={Ahn, Junwhan and Hong, Sungpack and Yoo, Sungjoo and Mutlu, Onur and Choi, Kiyoung},
  howpublished={Retrospective Issue for ISCA-50},
  year={2023}
}

@misc{mutlu2023rowhammerretrospective,
  title={{Retrospective: Flipping Bits in Memory Without Accessing Them: An Experimental Study of DRAM Disturbance Errors}},
  author={Onur Mutlu},
  howpublished={Retrospective Issue for ISCA-50},
  year={2023}
}

@misc{mutlu2023experimentalretrospective,
  title={{Retrospective: An Experimental Study of Data Retention Behavior in Modern DRAM Devices: Implications for Retention Time Profiling Mechanisms}},
  author={Onur Mutlu},
  howpublished={Retrospective Issue for ISCA-50},
  year={2023}
}

@misc{mutlu2023raidrretrospective,
  title={{Retrospective: RAIDR: Retention-Aware Intelligent DRAM Refresh}},
  author={Onur Mutlu},
  howpublished={Retrospective Issue for ISCA-50},
  year={2023}
}

@misc{mutlu2023selfretrospective,
  title={{Retrospective: Self-optimizing Memory Controllers: A Reinforcement Learning Approach}},
  author={Jos{\'e} F. Mart{\'\i}nez and Engin Ipek and Onur Mutlu and Rich Caruana},
  howpublished={Retrospective Issue for ISCA-50},
  year={2023}
}

@misc{yuksel2026memory,
  title={{Memory-Centric Computing: Security Benefits and Challenges of Processing-in-DRAM}},
  author={Ismail Emir Yuksel and F. Nisa Bostanci and Ataberk Olgun and Onur Mutlu},
  howpublished={The 7th Workshop on Memory-Centric Computing Systems colocated with ICS},
  year={2026}
}

@misc{luo2026ramulator2.1,
  title={{Ramulator 2.1: A Composable Memory System Simulator for Modern DRAM Systems}},
  author={Haocong Luo and F. Nisa Bostancı and Ataberk Olgun and Maria Makeenkova and Ziad Malik and Ipek Akdeniz and Onur Mutlu},
  howpublished={The 3rd Tutorial on Ramulator and DRAM Bender colocated with ICS},
  year={2026}
}

@misc{olgun2026drambender,
  title={{A Modern Large-Scale Memory Characterization Laboratory}},
  author={Ataberk Olgun and Haocong Luo and Ismail Emir Yuksel and F. Nisa Bostanci and A. Giray Yaglikci and Onur Mutlu},
  howpublished={The 3rd Tutorial on Ramulator and DRAM Bender colocated with ICS},
  year={2026}
}

@incollection{mutlu2015main,
  title={{Main Memory Scaling: Challenges and Solution Directions}},
  author={Mutlu, Onur},
  booktitle={More than Moore Technologies for Next Generation Computer Design},
  year={2015},
  publisher={Springer}
}

@article{gomez2021benchmarking,
  title={{Benchmarking a New Paradigm: An Experimental Analysis of a Real Processing-In-Memory Architecture}},
  author={G{\'o}mez-Luna, Juan and Hajj, Izzat El and Fernandez, Ivan and Giannoula, Christina and Oliveira, Geraldo F and Mutlu, Onur},
  journal={IEEE Access},
  year={2022}
}

@inproceedings{kim2014flipping,
title        = {{Flipping Bits in Memory Without Accessing Them: An Experimental Study of DRAM Disturbance Errors}},
author       = {Kim, Yoongu and Daly, Ross and Kim, Jeremie and Fallin, Chris and Lee, Ji Hye and Lee, Donghyuk and Wilkerson, Chris and Lai, Konrad and Mutlu, Onur},
year         = 2014,
booktitle    = {ISCA}
}

@inproceedings{kakolyris2026columnkeeper,
author = {Kakolyris, Andreas Kosmas and Bostanci, F. Nisa and Olgun, Ataberk and Yuksel, Ismail Emir and Songara, Harsh and Sgouras, Konstantinos Marios and Baser, Umut and Kanellopoulos, Konstantinos and Yaglikci, A. Giray and Mutlu, Onur},
title = {{ColumnKeeper: Efficient Solutions for Mitigating ColumnDisturb in DRAM-based Systems}},
booktitle = {ISCA},
year = {2026}
}

@article{ghose2019processing,
  title={{Processing-in-Memory: A Workload-driven Perspective}},
  author={Ghose, Saugata and Boroumand, Amirali and Kim, Jeremie S and G{\'o}mez-Luna, Juan and Mutlu, Onur},
  journal={IBM JRD},
  year={2019}
}

\end{document}